\documentclass[sigconf]{acmart}

\def\BibTeX{{\rm B\kern-.05em{\sc i\kern-.025em b}\kern-.08emT\kern-.1667em\lower.7ex\hbox{E}\kern-.125emX}}
    
\acmDOI{XXX}

\acmISBN{XXX}

\acmConference[CIKM '19]{ACM CIKM conference}{November 2019}{Beijing, China}
\begin{document}

%
% The "title" command has an optional parameter, allowing the author to define a "short title" to be used in page headers.
\title{Large-Scale Visual Search with Binary Distributed Graph at Alibaba}

%
% The "author" command and its associated commands are used to define the authors and their affiliations.
% Of note is the shared affiliation of the first two authors, and the "authornote" and "authornotemark" commands
% used to denote shared contribution to the research.
\author{Kang Zhao,Pan Pan,Yun Zheng,Yanhao Zhang,Changxu Wang,Yingya Zhang,Yinghui Xu,Rong Jin}
%\author{XXXX}
\affiliation{%
  \institution{Machine Intelligence Technology Lab, Alibaba Group}}
\email{zhaokang.zk,panpan.pp,zhengyun.zy,yanhao.zyh,changxu.wcx,yingya.zyy,renji.xyh,jinrong.jr@alibaba-inc.com}
%\email{XXXX@alibaba-inc.com}

%
% By default, the full list of authors will be used in the page headers. Often, this list is too long, and will overlap
% other information printed in the page headers. This command allows the author to define a more concise list
% of authors' names for this purpose.
\renewcommand{\shortauthors}{Kang Zhao, et al.}

%
% The abstract is a short summary of the work to be presented in the article.
\begin{abstract}
Graph-based approximate nearest neighbor search has attracted more and more attentions due to its online search advantages. Numbers of methods studying the enhancement of speed and recall have been put forward. However, few of them focus on the efficiency and scale of offline graph-construction. For a deployed visual search system with several billions of online images in total, building a billion-scale offline graph in hours is essential, which is almost unachievable by most existing methods. 

In this paper, we propose a novel algorithm called {\it Binary Distributed Graph} to solve this problem. Specifically, we combine binary codes with graph structure to speedup online and offline procedures, and achieve comparable performance with the ones in real-value based scenarios by recalling more binary candidates. Furthermore, the graph-construction is optimized to completely distributed implementation, which significantly accelerates the offline process and gets rid of the limitation of memory and disk within a single machine. Experimental comparisons on Alibaba Commodity Data Set (more than three billion images) show that the proposed method outperforms the state-of-the-art with respect to the online/offline trade-off.
\end{abstract} 

%\begin{CCSXML}
%<ccs2012>
%  <concept_id>10010520.10010553.10010562</concept_id>
%  <concept_desc>Computer systems organization~Embedded systems</concept_desc>
%  <concept_significance>500</concept_significance>
%  <concept_id>10010520.10010575.10010755</concept_id>
%  <concept_desc>Computer systems organization~Redundancy</concept_desc>
%  <concept_significance>300</concept_significance>
%  <concept_id>10010520.10010553.10010554</concept_id>
%  <concept_desc>Computer systems organization~Robotics</concept_desc>
%  <concept_significance>100</concept_significance>
%  <concept_id>10003033.10003083.10003095</concept_id>
%  <concept_desc>Networks~Network reliability</concept_desc>
%  <concept_significance>100</concept_significance>
%</ccs2012>
%\end{CCSXML}
%
%\ccsdesc[500]{Computer systems organization~Embedded systems}

%
% Keywords. The author(s) should pick words that accurately describe the work being
% presented. Separate the keywords with commas.

%
% A "teaser" image appears between the author and affiliation information and the body 
% of the document, and typically spans the page. 

\begin{teaserfigure}
  \includegraphics[width=\textwidth]{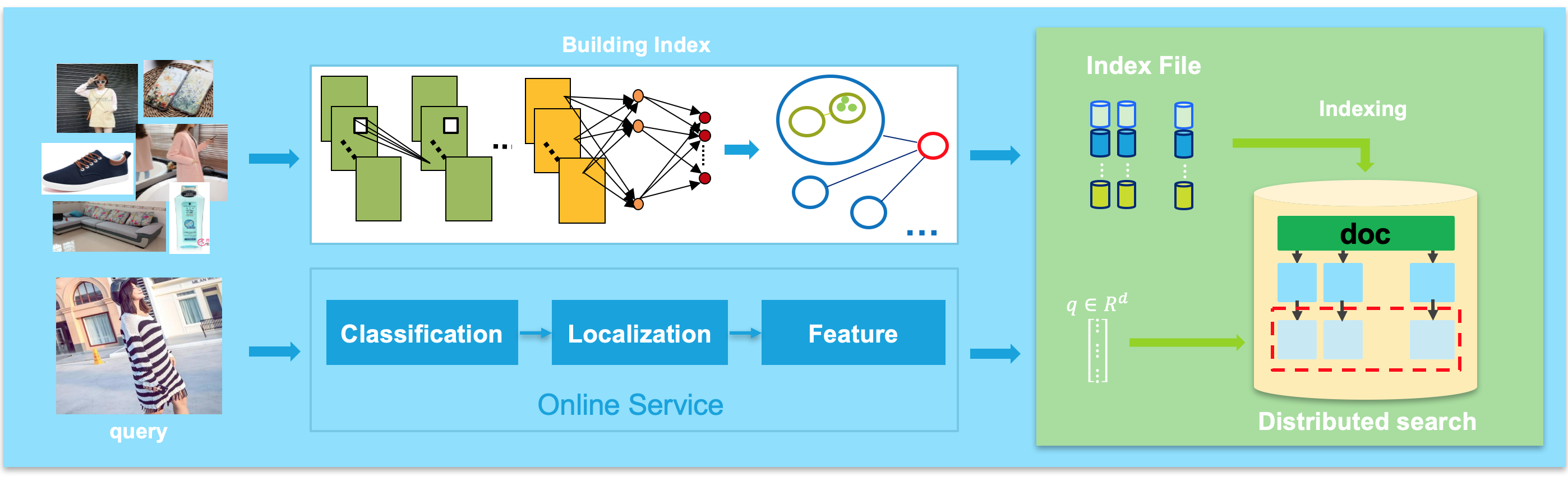}
  \caption{The total pipeline of "Pailitao" (visual search application at Alibaba).}
  \label{fig:pailitao}
\end{teaserfigure}

%
% This command processes the author and affiliation and title information and builds
% the first part of the formatted document.
\maketitle

\section{Introduction}
Recently, approximate nearest neighbor search (ANNS) has attracted increasingly attentions in many applications, like data mining, information retrieval and pattern recognition, to handle the explosive growth of the data on the internet. In the last decades, a lot of methods have been proposed from different aspects, such as hashing-based algorithms \cite{gui2018r,cao2016deep,weiss2008spectral,gong2011iterative,datar2004locality,zhao2014locality,zhao2014locality2,kong2012isotropic,xu2013harmonious,gao2015selective}, tree-based algorithms \cite{silpa2008optimised,beckmann1990r,bentley1975multidimensional,jagadish2005idistance,babenko2017product}, quantization-based algorithms \cite{jegou2011product,ge2013optimized,babenko2016efficient,babenko2012inverted,matsui2015pqtable,heo2014distance} and graph-based algorithms \cite{wang2013fast,harwood2016fanng,malkov2018efficient,dong2011efficient,malkov2014approximate,wang2012query}. Among them, graph-based approaches show superior performance than others in terms of search efficiency and recall, which has been demonstrated in many literatures \cite{douze2018link,harwood2016fanng,malkov2018efficient}. One important reason derives from the pre-calculated graph structure, which makes the traversal rapidly converge to the nearest neighbors of a given query, at the expensive cost of offline procedure.

At Alibaba, a typical scenario of ANNS is visual search. It has been studied for many years, and gives birth to a successful intelligence E-commercial application named ``\verb|Pailitao|''. ``\verb|Pailitao|'' is an innovative image search product based on deep learning and large scale ANNS algorithms, which provides the function of ``search by images'' via retrieving the photos taken by users. Figure \ref{fig:pailitao} shows its total pipeline. Unfortunately, the existing graph-based ANNS methods can hardly be used in ``\verb|Pailitao|'' with two main reasons: 1) The data size of ``\verb|Pailitao|'' is up to several billions in total. Limited by the CPU memory size, it's impractical to load all the data into a single machine to build the graph; 2) Constructing a billion-scale and high-quality offline graph is very time-consuming for the most existing graph-based methods, usually more than 20 hours that is unaccepted in ``\verb|Pailitao|''.

Apparently the offline graph construction is a key factor to make the graph-based ANNS methods applied in practice. Compared with the exact k nearest-neighbor (k-NN) graphs (a data connecting to its k nearest neighbors), the approximate k-NN graphs are preferred because of feasibility, and many efforts have been taken to optimize building it with higher scalability and efficiency. One of the possible solutions is based on ANNS algorithms \cite{uno2009efficient,zhao2018k,malkov2018efficient,malkov2014approximate}. We can regard each data as a query and find its approximate k-NNs by searching the indexing structure that is constructed in advance. But it's still very time-consuming if we want to build a billion-scale graph, because the ANNS process will be executed a billion times, no matter we make the search distributed or not.

The divide-and-conquer methodology is another potential approach \cite{wang2012scalable,wang2013fast}, which usually consists of two stages: divide-and-conquer and neighborhood propagation. In the first stage, one recursively partitions the data set into small subsets to build many subgraphs, and repeats it several times for finding more true neighbors. Then in the second stage, refinement strategies (such as local join or graph-based ANNS) are conducted for every data point to expand its neighborhood on the base approximate neighborhood graph obtained above. Due to the repetition of the first stage and the infeasible distributed implement of refinement schemes, it also suffers from high time complexity and memory limitation.

In this paper, we propose a novel graph-based ANNS algorithm named {\it Binary Distributed Graph} (BDH) to solve the above problems. The main contributions of our work are outlined as follows:
\begin{itemize}
\item We integrate binary codes with graph structure, which offers the efficient comparison in both online search and offline building procedure with Hamming distance, and attains comparable performance with the ones in real-value based scenarios by recalling more binary candidates.
\item A fully distributed graph-construction process is proposed, including single-pass divide-and-conquer algorithm based on binary clustering and the optimization of neighborhood propagation for distributed deployment, which can be easily implemented under MapReduce framework without the limitation of memory and disk within a single machine.
\item Experiments on the billion-scale Alibaba Commodity Data Set show that our algorithm outperforms the state-of-the-art with respect to the online/offline trade-off. What's more, we successfully make the graph-based ANNS method deployed in ``\verb|Pailitao|'' come true.
\end{itemize}

The rest of this paper is organized as follows. Section 2 reviews recent literatures on graph-based ANNS algorithms. We present our proposed approach in Section 3. Experimental results and analysis are demonstrated in Section 4. Finally, we make conclusions in Section 5.

\section{Related Work}
Generally speaking, graph-based ANNS methods comprise two parts: offline graph building and online search. Different algorithms usually construct the graphs with various properties which result in different search performance. In this section, we present a few popular approaches, especially, with their offline graph-construction process.

\subsection{$k$-Nearest Neighbor Graph}
It represents a series of methods, which refers to Kgraph \cite{dong2011efficient} here. Wei Dong {\it et al.} propose an efficient algorithm called NN-Descent (NND) for approximate k-NN graph construction, following simple principle: ``{\it a neighbor of a neighbor is also likely to be a neighbor}'' (we call it {\it neighbor principle} for short). It's noted that NND is suitable for large-scale applications where the data is located throughout the network, since it does not depend on any shared global index, and only needs local search. Nevertheless, when implementing it under a distributed framework like Spark, we find it is not fast as expected. As an iterative-oriented algorithm, NND needs to exchange many pair-data between different nodes within each iteration, which is not friendly to distributed design, and even makes it perform worser than a single machine.

\subsection{Navigable Small World Graph}
Navigable small world graph (NSW) \cite{malkov2014approximate} builds a navigable small world graph with a variation of greedy search to solve the approximate k-nearest neighbor search problem, where nodes corresponding to data point, edges to relations between them. Interestingly, it provides a very simple way to construct the NSW graph: inserting elements in random order and connecting them up to $M$ closest neighbors from the previously built graph. Although insertion and $k$-NN search can be done in parallel, we need to conduct the search $N$ times in total, which takes very long for a billion-scale graph building.

\subsection{Hierarchical Navigable Small World Graph}
As the most accomplished version of graph-based approaches, hierarchical navigable small world graph (HNSW) \cite{malkov2018efficient} selects a series of nested subsets of database vectors to construct the hierarchical structure. The sizes of the layers decrease with logarithmic scaling, beginning with the base layer that contains the whole dataset, ending up with the first layer of just one point. Besides, there is a neighborhood graph in each of these layers. The top-to-down search finding the nearest neighbor (marked as $A$) of the query within one layer goes down the next layer to continue the search from $A$, except for the base layer, which performs a hill-climbing strategy.

As Yuri Malkov {\it et al.} claim in \cite{malkov2018efficient}, the apparent shortcoming of HNSW lies in the loss of the possibility of distributed search in the graph-construction process, due to the complex hierarchical structure. Even if we succeed in distributing it finally, it also suffer from the same issue as NSW.

\begin{figure*}[t]
  \includegraphics[width=1.0\textwidth]{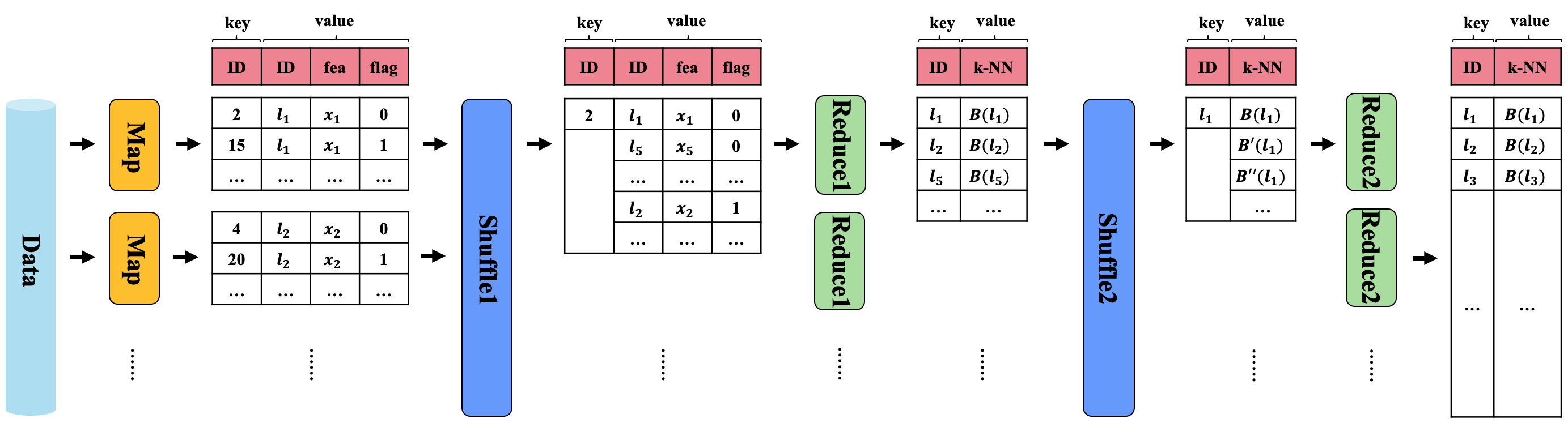}
  \caption{The data flow of single-pass divide-and-conquer in MapReduce pipeline.}
  \label{fig:dataflowdac}
\end{figure*}

\subsection{Fast Approximate Nearest Neighbor Graph}
\cite{harwood2016fanng} presents the design of an ideal graph structure by pruning the edges of each node to approximate the intrinsic dimensionality of the local manifold-like structure, which leads to efficient searching. In order to get high recall for query locating anywhere, fast approximate nearest neighbor graph (FANNG) improves the search algorithm with backtracking. Specifically, it goes back to the second closest vertex that is kept in a priority-queue if it fails to find new closer one.

However, the ideal graph construction has a high time complexity $O(n^2\log(n))$ and is extremely expensive for a large graph building. Despite being accelerated with an efficient strategy \cite{harwood2016fanng}, it still needs to call the traverse-add function $50N$ times, where $N$ is the dataset size. After that, a second stage is required to further boost the graph. In addition, no distributed solution is provided. 

\section{Binary Distributed Graph}

This section presents the formulation of our {\it Binary Distributed Graph} (BDG) algorithm. First, we introduce the binary codes generation. Then, a completely distributed graph-construction process is presented, including two main parts: single-pass divide-and-conquer algorithm and distributed neighborhood propagation. Finally, we show the search procedure and some algorithm details. The notations are given below to facilitate our description.

Consider a dataset with $n$ samples $\{l_i, \mathbf{x_i}\}_{i=1}^n$, where $\mathbf{x}\in\mathrm{R}^{d}$ and $l$ represents the unique label for each data. Let $B(l_i)$ be the approximate $k$-NN label set of $i$-th data, and $R(l_i)$ be the reverse one that is defined as $R(l_i)=\{l|l_i \in B(l)\}$. Denote $d : \mathrm{R}^{d} \times \mathrm{R}^{d} \rightarrow \mathrm{R}$ the similarity measure, we mainly consider the case $d=l_2$ and $d=l_h$ (Hamming Distance), which is commonly used in computer vision applications, as well as ``\verb|Pailitao|''.

\subsection{Generating Binary Codes}
As described in \cite{zhang2018visual}, we first obtain the real-value representations by CNN embeddings based on user click data. Then, the features are mapped to the Hamming space to preserve their original locality structures, which can be accomplished by a lot of ways, such as ITQ \cite{gong2011iterative}, SH \cite{weiss2008spectral} and the hashing techniques based on deep learning \cite{cao2016deep,gui2018r}. In practice, we adopt LPH \cite{zhao2014locality} to learn a compact $d$-dimensional binary code $\mathbf{y} \in \{0, 1\}^d$.

Note that the usage of our binary codes is different from that in Polysemous Code \cite{douze2016polysemous}, where it is used for filtering the indexed vectors. And the indexing algorithm in \cite{douze2016polysemous} is designed for real value feature in essence. By contrast, we convert the query and doc vectors to bits from the beginning to significantly promote the efficiency both in offline and online processes.

\subsection{Single-Pass Divide-And-Conquer}
Similarly, we take advantage of the divide-and-conquer methodology to build a base approximate neighborhood graph. Unlike others recursively splitting the space, we exploit a flat clustering method like $k$-means, which can be distributed very easily. In order to make full use of efficient Hamming Distance calculation, we require the centers to be binary too. Given $m$ centers, then we have the following formulation: 
\begin{eqnarray}
J(r, \mathbf{c})=\min \sum_i^n \sum_j^m r_{ij} \| \mathbf{y}_i - \mathbf{c}_j \|_2^2 \\ \nonumber
\mathrm{subject\ to}: \mathbf{c}_j \in \{0, 1\}^d
\end{eqnarray}
where $\mathbf{c}_j$ is the $j$-th center, and $r_{ij}=1$ if the $i$-th data belongs to $\mathbf{c}_j$ (0 for otherwise). It is equivalent to Bk-means \cite{gong2015web} in form, then a similar but not identical updating strategy is put forward : 
\\\textbf{Assigning Step: fix c and optimize r.} We assign each data to its nearest center, just like $k$-means. Instead of building a multi-index hash table for millions of clusters \cite{gong2015web}, we adopt exhaustive comparisons among the centers, considering our limited number of centers and the simple linear computation is easier to be distributed. 
\\\textbf{Updating Step: fix r and optimize c.} Assuming there are $p$ data points in center $\mathbf{c}_j$, then $\mathbf{c}_j$ can be updated by: 
\begin{eqnarray}
\mathbf{c}_j = sgn(\sum_i^p \mathbf{x}_i)
\end{eqnarray}

\begin{figure}[h]
  \includegraphics[width=0.5\textwidth]{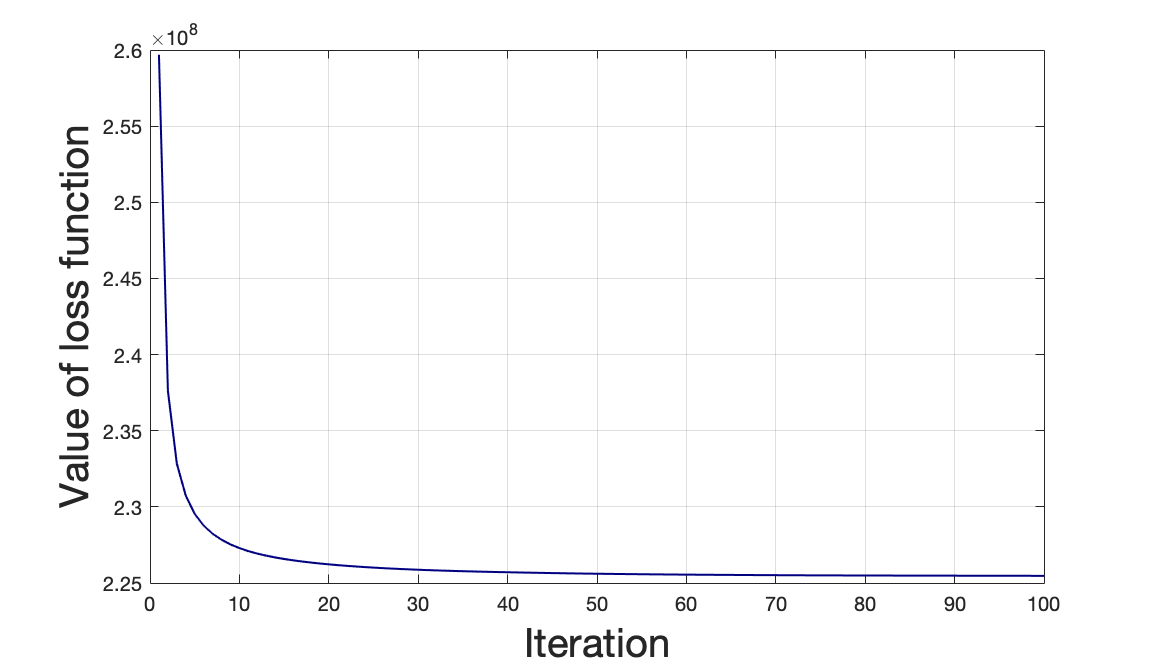}
  \caption{The values of loss function in Bk-means.}
  \label{fig:objectivevalue}
\end{figure}

As a clustering method, Bk-means is proposed to cluster huge number of photos on a single machine \cite{gong2015web}. Consequently, it involves the total dataset and long iterations to make the loss value small enough, leading it quite different from ours. What we need is the binary centers, so we just down-sample a fraction of data to conduct Bk-means. Figure \ref{fig:objectivevalue} shows that the loss function value descend very quickly at the beginning, but decrease slowly after several iterations. We set the number of iterations less than 10, which offers a good trade-off between time and accuracy. In addition, we implement our Bk-means in an iterative-oriented distributed framework at Alibaba, instead of a single machine.

\begin{figure}[b]
  \includegraphics[width=0.48\textwidth]{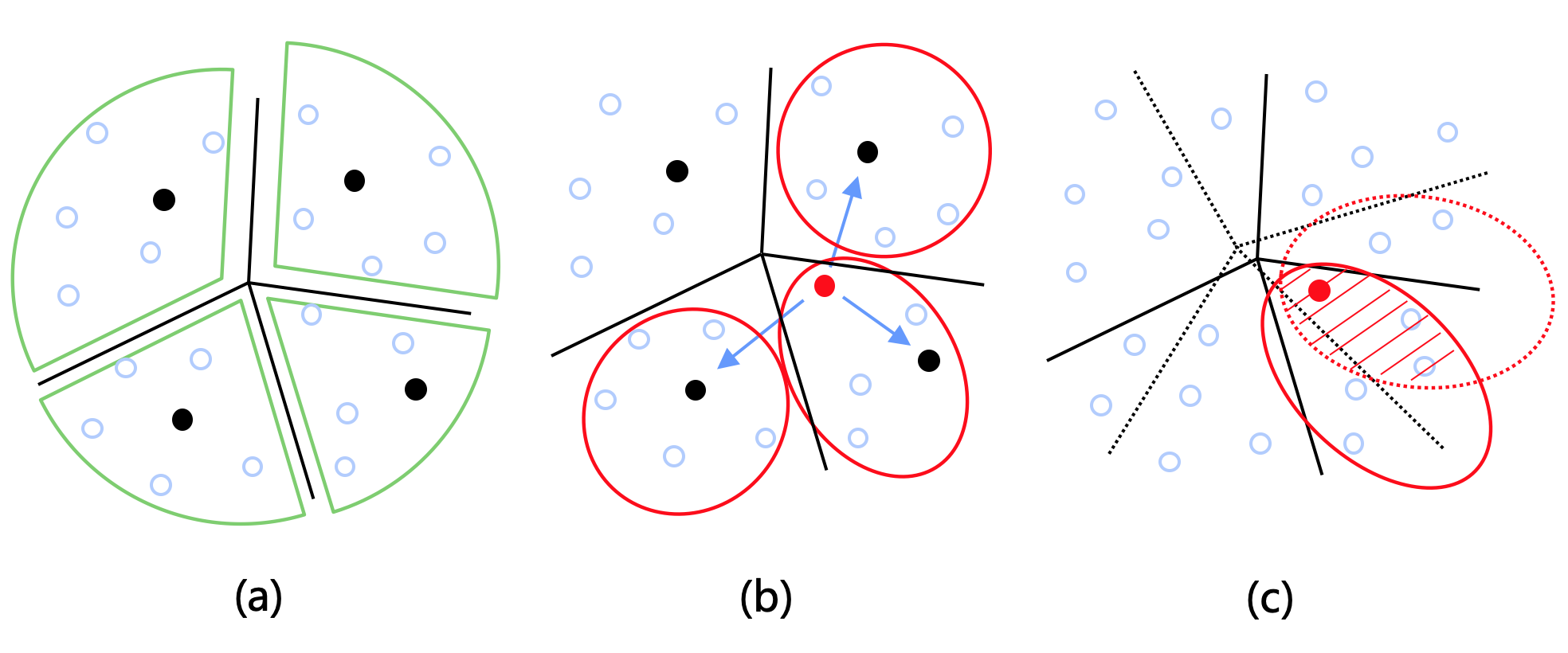}
  \caption{We compare our method with the multiple random one. (a) maps data to its nearest center; (2) maps data to several nearest centers; (3) shows the multiple random one.}
  \label{fig:subgraphs}
\end{figure}

If we map each data point to its nearest center, the base approximate neighborhood graph generated will comprise $m$ isolated subgraphs, without connecting their neighborhoods lying in different subgraphs as shown in Figure \ref{fig:subgraphs}(a). Hence, we find $t$ nearest centers for every data to get more neighbors within a single divide-and-conquer process. Compared with multiple random process \cite{wang2012scalable}, our single-pass method can avoid redundant computations as illustrated in Figure \ref{fig:subgraphs}(b)(c). The overlaps between any two consecutive partitions lead to less and less benefits as the number of random division increases. What's worse, multiple random divide-and-conquer will repeat the process many times, introducing extra non-ignorable computational cost. Moreover, considering each center having different number of data, we fix the sum of points (donated as $coarse\_num$) included in $t$ nearest centers to make the computation not biased.

The single-pass divide-and-conquer process is executed in MapReduce of three phases: Map + Reduce1 + Reduce2. Figure \ref{fig:dataflowdac} shows the data flow in details, where $B'(l_i)$ means another $k$-NN group of data $l_i$, so does $B''(l_i)$. The Map function gets the input data from the database and compute their $m$ nearest centers. Then it outputs the key-values records, where key is the index of the centers and value contains data feature and label. An extra flag (0/1) is required to mark whether the center that the record belongs to is nearest. After the Shuffle stage, the records with the same key will be merged together. In the Reduce1 phase, we treat all the input records as queries to search the records with $flag=0$. The output key is the data label with its $k$-NN as the value. We merge all the records having the same data label in the second Shuffle, and sort out the final $k$-NN candidates in the function of Reduce2.

Taking account of the low data transmission efficiency of MapReduce, we limit the number of clusters, which not only increases the size of each center but also reduces the number of output records generated in Map stage, to make the calculation intensive.

\begin{figure}[t]
  \includegraphics[width=0.48\textwidth]{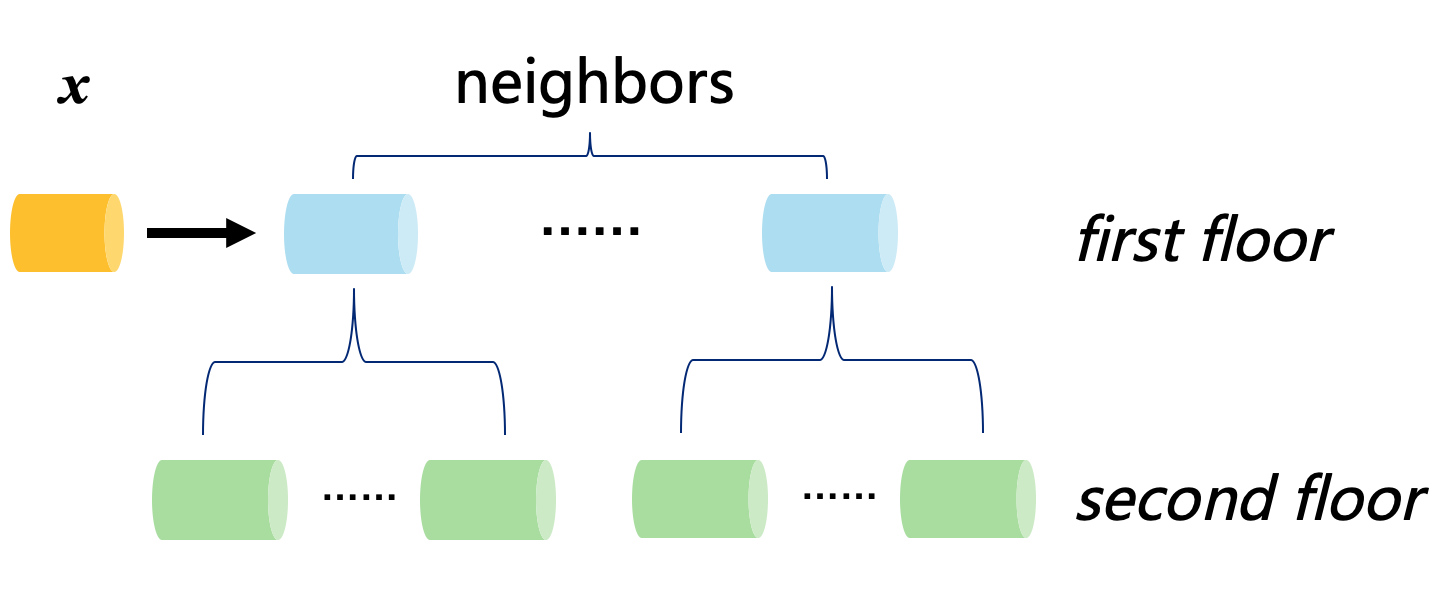}
  \caption{We expand all the neighbors of data x. First floor means the neighbors with depth = 0; Second floor means depth = 1.}
  \label{fig:doublefloor}
\end{figure}

\subsection{Distributed Neighborhood Propagation}

\begin{figure}
  \includegraphics[width=0.39\textwidth]{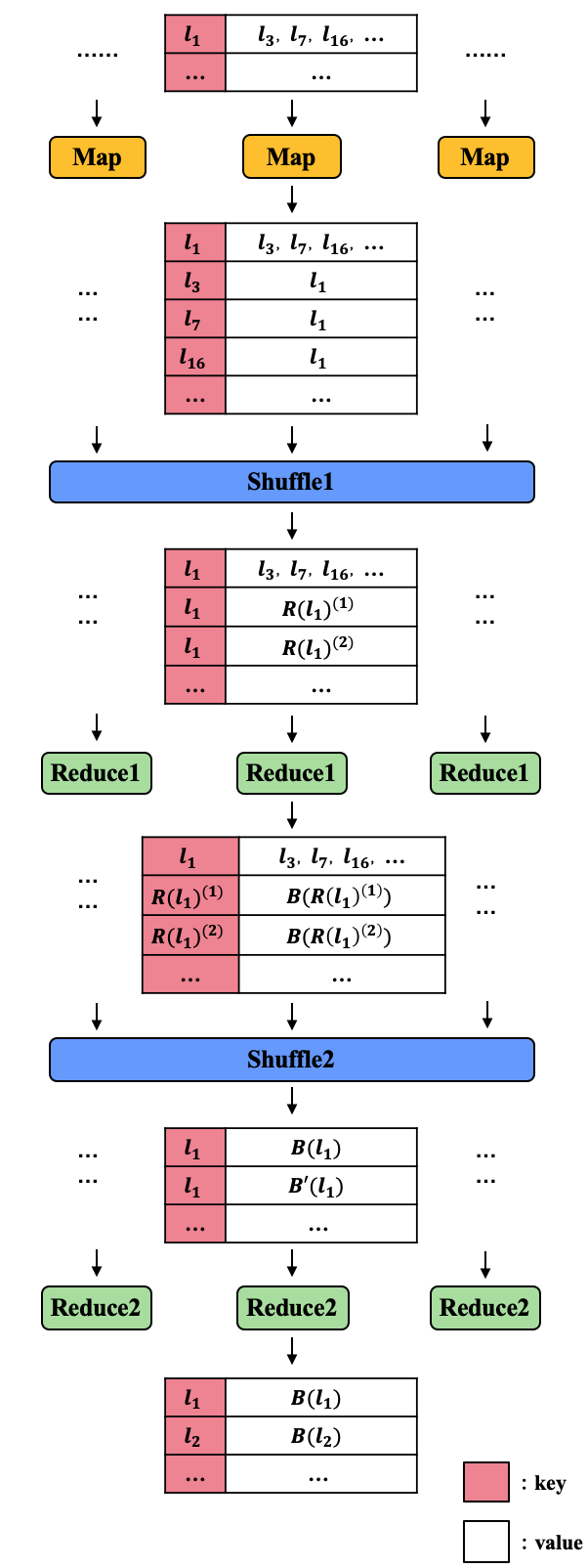}
  \caption{The data flow of distributed neighborhood propagation in MapReduce pipeline.}
  \label{fig:dataflowdnp}
\end{figure}

Same as NND, neighborhood propagation also bases upon the {\it neighbor principle}. But the existing methods are designed for single machine. Jing Wang {\it et al.} in \cite{wang2012scalable}, taking the neighborhood in the base graph as starting points, execute the graph-based ANNS to refine the neighbors, which is difficult to be performed for all the points in a distributed and efficient pipeline. Besides, the local join strategy in \cite{wang2013fast} exchanges the information between the neighbor pairs, and it will also encounter the same problem as NND.

Why the above schemes can't be distributed easily? To some extent, they are complicated. We try to simplify the propagation. Thanks to the high quality of the base graph, replacing depth-first search with breadth-first one and increasing depth by one every time becomes a sound approach. As displayed in Figure \ref{fig:doublefloor}, we expand all the first floor neighbors of data $\mathbf{x}$ in a flat way, then compare $\mathbf{x}$ with all the neighbors of the second floor. This design is simple yet friendly to distributed framework and can be repeated several times for more true neighbors.

We show the data flow in Figure \ref{fig:dataflowdnp}. Map function gets records like $[l_1, \{l_3, l_7,l_{16},...\}]$ as input, and maps it out as $[l_3, l_1], [l_7, l_1], etc$. After the Shuffle stage, the key $=l_1$ records will be merged, including its original and reverse neighbors, where $R(l_1)^{(1)}$ means one reverse neighbor of $l_1$, the same to $R(l_1)^{(2)}$. Next, we calculate each reverse neighbors with all the base neighborhood in the Reduce1 function. With the second Shuffle, all the neighbors (in the second floor) of key $=l_1$ will be together and be merge-sort out after the Reduce2.

\subsection{Offline Infrastructure}

Inspired by FANNG \cite{harwood2016fanng}, we also prune our graphs to save the memory and boost the search, and add it into the distributed pipeline.

In ``\verb|Pailitao|'', a multi-replications and multi-shards index engine architecture is adopted, which is not only good scalability, but also robust fault-tolerant. Multi-shards mean multiple machines are deployed to store the total data (including vectors, index structures and other storage), each shard storing only a subset. As a result, it is necessary to support the multi-shards graphs building during the offline process. 

\begin{figure}[t]
  \includegraphics[width=0.48\textwidth]{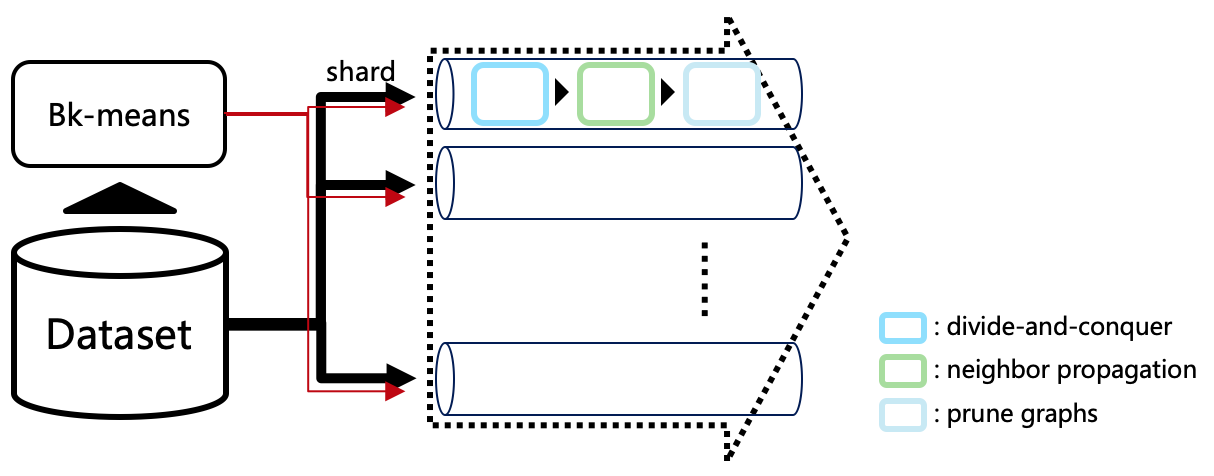}
  \caption{The offline infrastructure.}
  \label{fig:offlinepipeline}
\end{figure}

As shown in Figure \ref{fig:offlinepipeline}, we design the offline infrastructure to make it capable of (1) building single or multi-shards graphs as you wish; (2) building multi-shards graphs parallelly. Note that, the Bk-means is implemented only once before splitting the dataset, since the centers generated are not sensitive to different shards. 

\subsection{Search Procedure}

Owing to the limited representations of binary codes, many discriminative messages will be lost during the process of hashing, which will bring about low recall inevitably.

Fortunately, we find that if we recall more binary candidates than usual and rerank them with their real-value features, it will provide the comparable performance with the start-of-the-art that is in total real-value scenarios. Concretely, we rerank all the data in the final candidate pool whose size is larger than target result set. Recall will be improved at the cost of less than 1000 euclidean distance calculations.

Our implementation of hill-climbing is similar to the ones in the mainstream methods. Different from the hierarchical search of HNSW, we randomly sample some points, and compare them with the query for finding the nearest one, which is used as the entry point of the graph-based search. These sampled points can be regarded as ``long-link'' (corresponding to the high layers in HNSW), and the global $k$-NN graph structure we build offline (corresponding to the base layer) is regarded as ``short-link''. Experiments (see 4.4) present if we recall dozens of nearest neighbors (for example, top 60) instead of the nearest one, the ``short-link'' will play more important role than the ``long-link''. And this, from another perspective, explains the importance of offline global $k$-NN graph-construction.

\subsection{Algorithm Details}

\begin{figure*}[t]
  \includegraphics[width=1.0\textwidth]{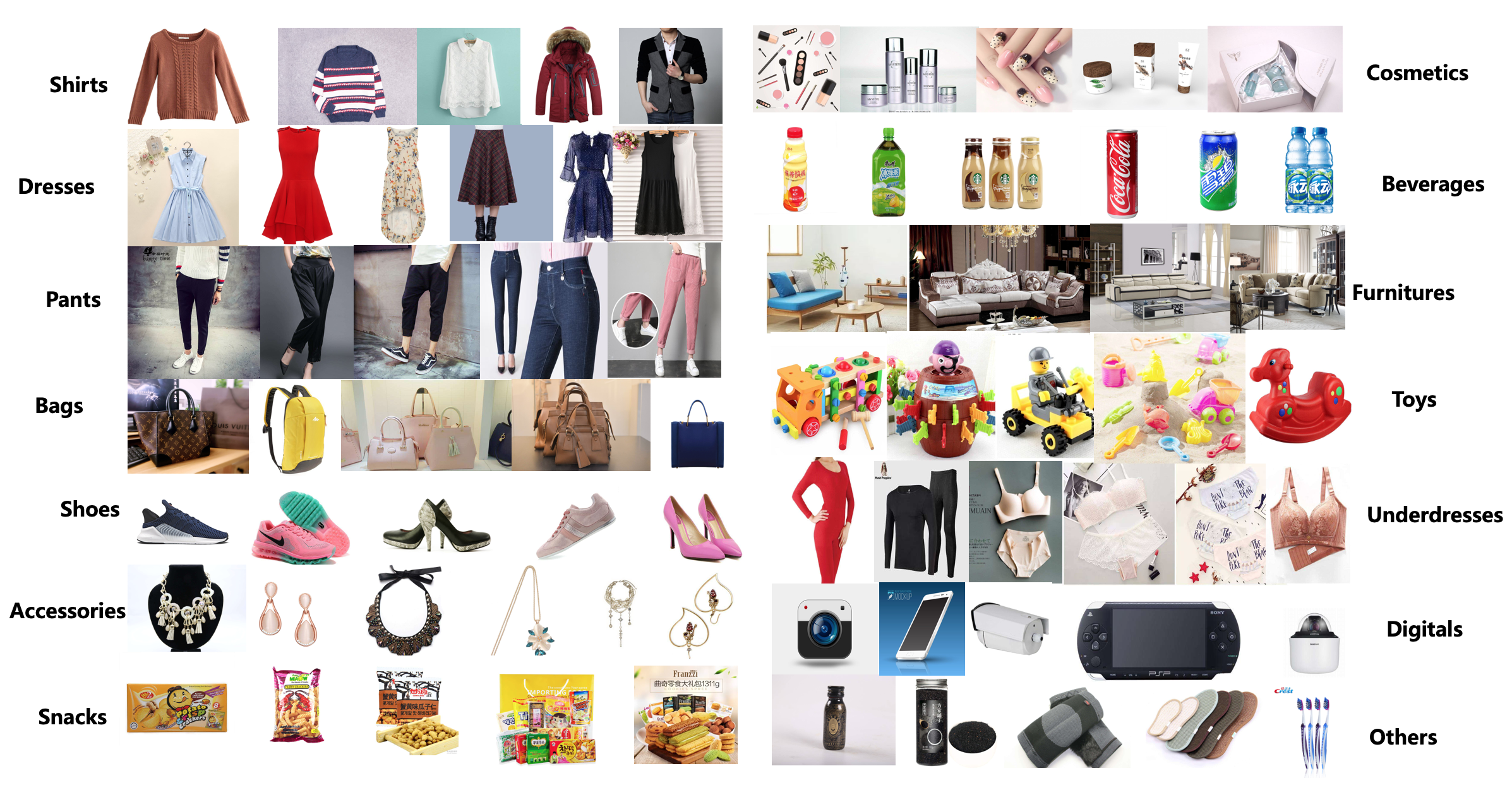}
  \caption{Some samples of the fourteen datasets.}
  \label{fig:dataphoto}
\end{figure*}

\begin{itemize}
\item[(1)] Data skew: In the single-pass divide-and-conquer stage, we randomly map the dataset to different Reduce1 nodes, and make each cluster contained in only one node. The data skew may happen if one has too many clusters than others. Therefore, we use a simple dynamic programming to shuffle the data, so that the total amount of data included on different nodes is close to each other.
\item[(2)] JNI: As we all know, Java is suitable for distributed computation. But it performs poorly than C++ in terms of efficiency, especially in the case of large numbers of distance calculations. In order to further improve distributed performance, we adopt JNI technique to speedup Hamming distance computation.
\item[(3)] Propagation filter: In the execution of distributed neighborhood propagation, we find the Shuffle process occupies the most time, caused by its low efficiency of data transmission. Actually, the base graph generated has a good quality of neighborhoods, which means the majority of neighbors in the second floor are not true ones, which are not necessary for transmission. We filter the second floor with the max distance between one point and its neighbors in the first floor, accelerating Shuffle2 by more than 50\%.
\end{itemize}

\section{Experiments}

\subsection{Datasets}
We evaluate our BDG method on the fourteen commodity datasets of ``\verb|Pailitao|'': shirts, dresses, pants, bags, shoes, accessories, snacks, cosmetics, beverages, furnitures, toys, underdresses, digitals and others, some of them are displayed in Figure \ref{fig:dataphoto}. In our experiments, images are represented as 512-dimensional real-value features which are trained by datasets, and mapped into 512-bits. Then we sample a subset from every dataset to enrich the diversity of scale. More details are listed in Table \ref{tab:datasets}.

\subsection{Evaluation Protocols and Baselines}

The performance of offline and online processes are assessed respectively. For offline, we compare the building time of different methods, and show both single and multiple shards time for our BDG. Online search takes the brute-force results, based on real-value, as ground truth to calculate the recall of ANNS. More specifically, given a query, we use its real-value expression to obtain the $k$-NN by exhaustive comparisons. Then the top$N$ recall is computed as:
\begin{eqnarray}
recall(N)=\frac{|B_{anns}(l) \cap B_{linear}(l)|}{N}
\end{eqnarray}
where $B_{anns}(l)$ means the top$N$ result is recalled by the ANNS algorithm, $B_{linear}(l)$ by the linear search, and $|S|$ counts the number of set $S$. We also record the search time. All the algorithms of online search are implemented in C++, compiled with same option used in hnswlib, and ran on a Linux machine with Intel(R) Xeon(R) CPU E5-2682 v4 2.50GHz and 512 GB memory.

Considering the relevance to our approach and the fairness of comparison, we choose the following state-of-the-art methods, which all provide Open Source Code:
\begin{itemize}
\item KGraph\footnote{\url{https://github.com/aaalgo/kgraph}}: We perform the NND algorithm with its default settings and use prune1 function (prune2 will significantly increase the overhead of offline and is not recommended by the author). Compiler options are revised (as stated above) to make its search procedure faster;
\item NSW\footnote{\url{https://github.com/nmslib/nmslib}}: The file I/O of nmslib is modified to adapt to our data type. We set the NN number of the new add points 50, and the number of neighbors is limited to 50 too.
\item HNSW\footnote{\url{https://github.com/nmslib/hnswlib}}: The hnswlib is an optimized version provided by nmslib with better performance of HNSW, which is also proved in our experiments.
\item BDG: This is our method proposed in this paper. We limit the number of neighbors no more than 50, and use the following settings: $m=8192, coarse\_num=100000$ if there is no other declaration.
\end{itemize}

\begin{table}[t]\fontsize{9.0pt}{\baselineskip}\selectfont
\centering
\caption{\label{tab:datasets}Introductions of the datasets, including dimension, bits, the numbers of images and queries.}
\begin{tabular}{|l||c|c|c|c|}
\hline
Dataset & $d$ & bits & size of images & size of queries\\ \hline
Shirts   		& 512 & 512 & 300M & 10000 \\
Dresses 		& 512 & 512 & 84M & 10000 \\
Pants    		& 512 & 512 & 130M & 10000 \\
Bags   		& 512 & 512 & 110M & 10000 \\
Shoes	 	& 512 & 512 & 150M & 10000 \\
Accessories    	& 512 & 512 & 240M & 10000 \\
Snacks   		& 512 & 512 & 20M & 10000 \\
Cosmetics 	& 512 & 512 & 110M & 10000 \\
Beverages    	& 512 & 512 & 24M & 10000 \\
Furnitures   	& 512 & 512 & 59M & 10000 \\
Toys 			& 512 & 512 & 46M & 10000 \\
Underdresses 	& 512 & 512 & 25M & 10000 \\
Digitals 		& 512 & 512 & 27M & 10000 \\
Others    		& 512 & 512 & 1.5B & 10000 \\
\hline
\end{tabular}
\end{table}

\subsection{Offline Building Time}
In order to maximize the performance of a single machine, we construct the graph for HNSW, KGraph and NSW with forty threads, and use no more than 2K cores in ODPS MR (the MapReduce framework at Alibaba) to build BDG. 

As shown in Table \ref{tab:buildtime}, our BDG method is the fastest on all datasets and makes the billion-scale graph building no more than five hours, proving its efficiency and scalability. Note that when the data size is small (e.g. 50M), due to resources allocation and data transmission, the time of BDG doesn't change linearly. Besides, multi-shards costs a little bit more than single one because of data splitting. Limited by the memory of single machine, HNSW, KGraph and NSW all suffer from failure on condition of too large dataset. By contrast, our BDG is distributed designed, which will hold very large data scale. Put aside the memory issue, HNSW performs better than KGraph and NSW since it makes full use of graph-based ANNS to speedup the building process and is well optimized in codes. KGraph utilizes NND algorithm to avoid excessive distance calculation. However, because of long iterations and frequent memory read-write overhead for pair exchanging, it's inferior to HNSW. The slowest method is NSW, which is simple but ineffective, leading to almost 2 hours for a 20M dataset. Our BDG method not only takes advantage of large-scale distributed system, but also accelerates the procedure with Hamming distance. Consequently, the experimental results demonstrate that our binary distributed strategy makes sense. 

Thanks to the distributed system, if we use more cores to construct a billion-scale or larger graph, the building time will drop almost linearly.

\begin{table}[t]\fontsize{9.5pt}{\baselineskip}\selectfont
\centering
\caption{\label{tab:buildtime}The offline building time of the fourteen datasets.}
\begin{tabular}{|l||c|c|c|c|c|} \hline
 &\multicolumn{2}{c|}{\bf BDG} & {\bf HNSW} & {\bf KGraph} & {\bf NSW}\\\cline{1-6}
 \#shards & 1 & 20 & —— & —— & —— \\ \hline
Shirts   		& \textbf{1h}    & 1.1h   & 5h & {\color{red} $\times$} & {\color{red} $\times$} \\
Dresses 		& \textbf{0.2h} & 0.23h & 1.4h & 7.3h & {\color{red} $\times$}\\
Pants    		& \textbf{0.4h} & 0.44h & 2.1h & 11.3h & {\color{red} $\times$}\\
Bags   		& \textbf{0.3h} & 0.32h & 1.8h & 9.6h &{\color{red} $\times$}\\
Shoes	 	& \textbf{0.4h} & 0.45h & 2.5h & 13h & {\color{red} $\times$}\\
Accessories    	& \textbf{0.8h} & 0.87h & 4h & {\color{red} $\times$} & {\color{red} $\times$}\\
Snacks   		& \textbf{0.1h} & 0.1h   & 0.3h & 1.74h & 1.83h \\
Cosmetics 	& \textbf{0.3h} & 0.31h & 1.8h & 9.6h & {\color{red} $\times$}\\
Beverages    	& \textbf{0.1h} & 0.1h   & 0.4h & 2h & 2.2h\\
Furnitures   	& \textbf{0.2h} & 0.21h & 1h & 5.1h & 5.4h\\
Toys 			& \textbf{0.2h} & 0.2h   & 0.75h & 4h & 4.2h\\
Underdresses 	& \textbf{0.1h} & 0.11h & 0.4h & 2.1h & 2.28h\\
Digitals 		& \textbf{0.1h} & 0.12h & 0.45h & 2.2h & 2.47h\\
Others    		& \textbf{5h}	   & 5.5h   & {\color{red} $\times$} & {\color{red} $\times$} & {\color{red} $\times$}\\
\hline
\end{tabular}
\end{table}

\subsection{Online Search Performance}

\begin{figure}[t]
  \includegraphics[width=0.5\textwidth]{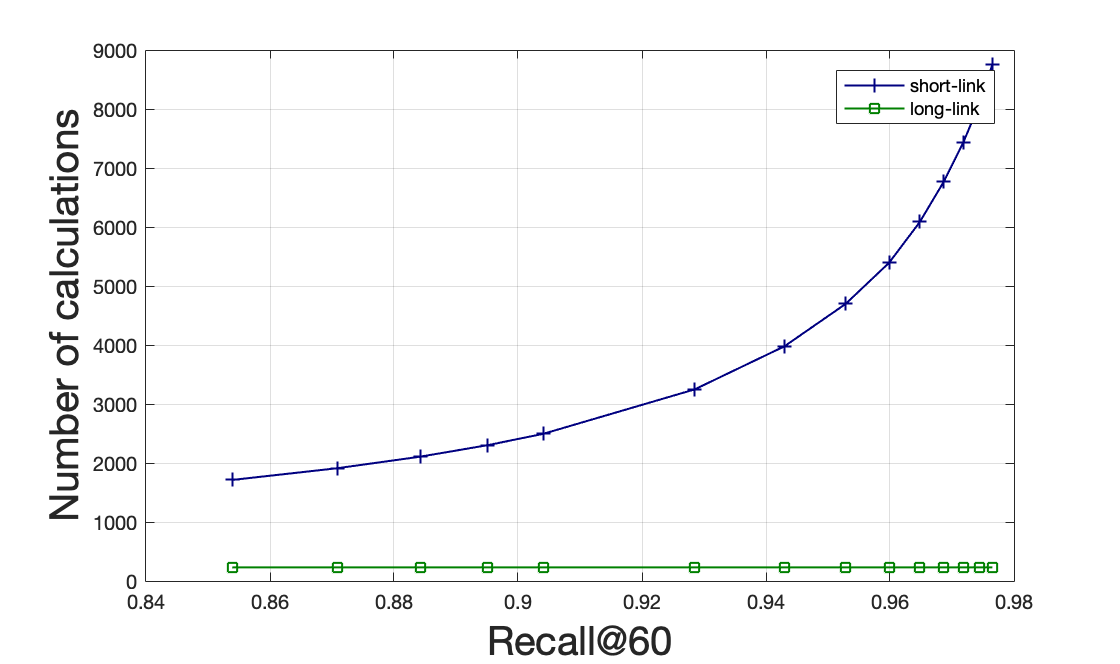}
  \caption{The number of distance computations in ``long-link'' and ``short-link''.}
  \label{fig:longshort}
\end{figure}

We first present the contributions between ``long-link'' and ``short-link'' in search process. Taking HNSW as an example, by adjusting the recall of top60, we record the number of distance computations in ``long-link'' and ``short-link'' respectively, as illustrated in Figure \ref{fig:longshort}. It can be seen that, as the recall increases, the calculation of ``short-link'' becomes more and more, but ``long-link'' is almost unchanged. Even at the lowest recall, the proportion of ``long-link'' to ``short-link'' is very low, let alone higher recall. The major role ``short-link'' plays in online search proves that our concentration on offline global $k$-NN graph building is reasonable.

\begin{figure*}[t]
  \includegraphics[width=1.0\textwidth]{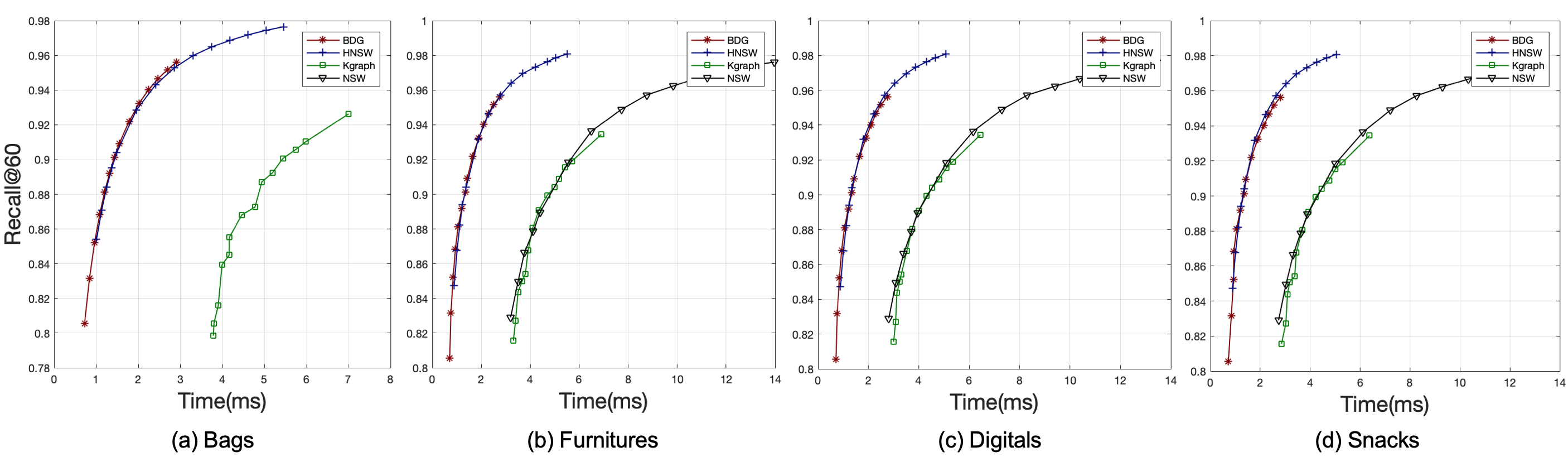}
  \caption{Recall-time curves on bags, furnitures, digitals and snacks.}
  \label{fig:recalltime}
\end{figure*}

In Figure \ref{fig:recalltime}, we choose four datasets to plot the recall-time curves of top60 with increasing the candidate pool size. It is clear that our BDG gets comparable performance with HNSW. The poor behaviors of KGraph and NSW have something to do with their codes, not well optimized as HNSW. On the other hand, the quality of their graphs makes it difficult to reach the nearest neighbor when given a query. By recalling more binary candidates and reranking them with real-value features, our BDG method obtains high recall as real-value ones. In consideration of both online and offline performance, our approach is superior to other state-of-the-art methods.

\subsection{Parameters Tuning}

\begin{figure}[t]
  \includegraphics[width=0.5\textwidth]{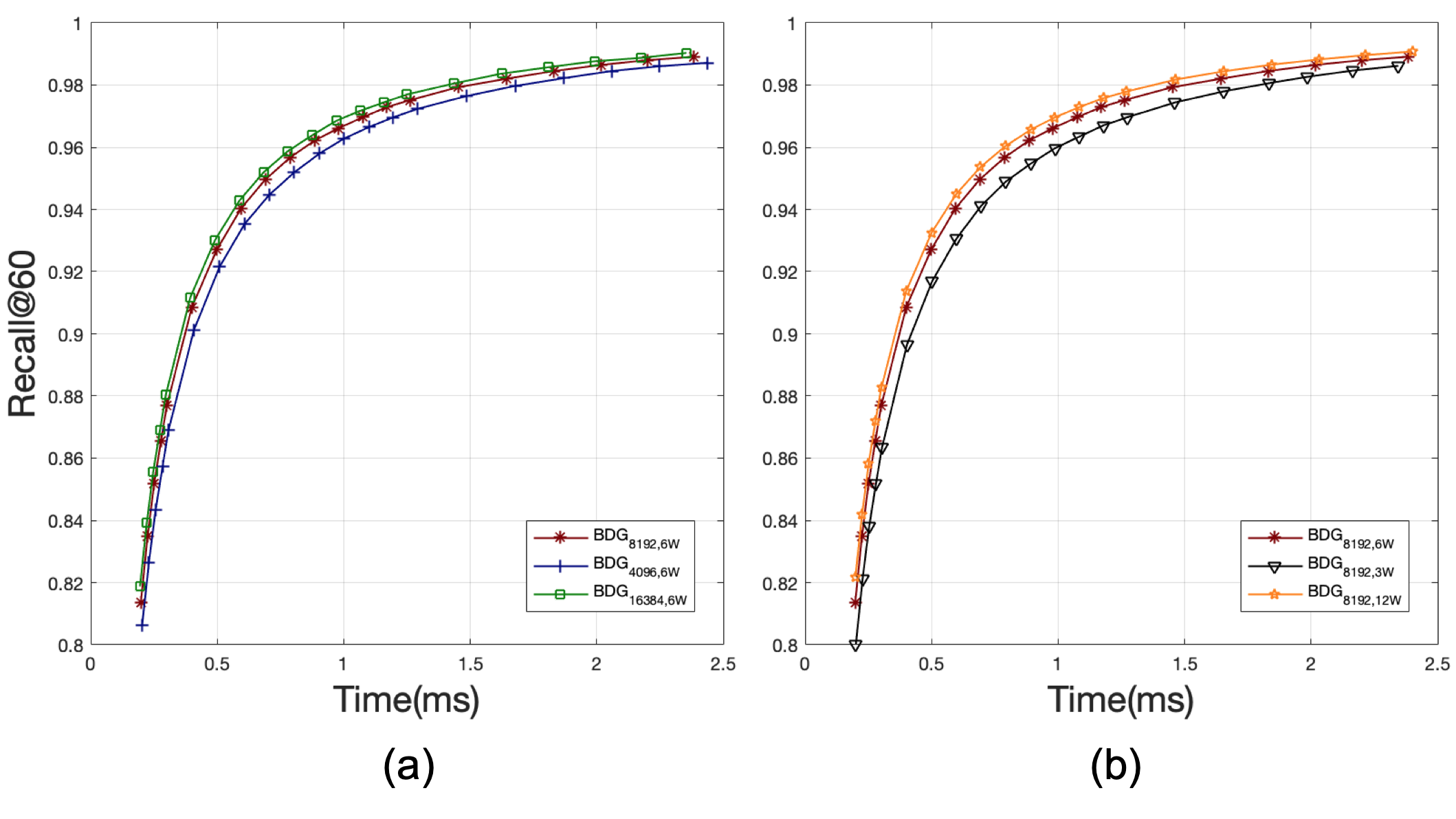}
  \caption{Recall-time curves with different (a) $m$ and (b) $coarse\_num$.}
  \label{fig:tuning}
\end{figure}

During the offline graph-construction of our BDG, there are many parameters that will affect the final quality of graphs, especially the number of clusters $m$ and the exhaustive comparison limitation $coarse\_num$ in the single-pass divide-and-conquer stage. To facilitate the exhibition of impact, we use binary linear search as ground truth instread.

Figure \ref{fig:tuning}(a) shows three curves with different $m$, and Figure \ref{fig:tuning}(b) with different $coarse\_num$. Obviously, the more center there are, the higher recall will be, so does $coarse\_num$. When we partition the space into more parts, the loss function value of Bk-means will be smaller, increasing the possibility of finding the nearest neighbor in the graph structure. But, as discussed earlier, we will not set $m$ too big for the sake of intensive computation. In addition, if we compare every node with more candidates, it will improve the whole quality of the base graph.

\subsection{Search in "Pailitao".}

\begin{table}[t]\fontsize{7.9pt}{\baselineskip}\selectfont
\centering
\caption{\label{tab:pailitao}Comparisons with the former search algorithm in "Pailitao".}
\begin{tabular}{|c||c|c|c|c|c||c|} \hline
 \#recall & top1 & top10 & top20 & top40 & top60 & time(ms)\\ \hline
Former   	& 98.47\% & 98.22\% & 98.03\% & 97.85\% & 97.70\% & 10\\
Now 		& \textbf{99.63\%} & \textbf{99.33\%} & \textbf{99.03\%} & \textbf{98.58\%} & \textbf{98.18\%} & \textbf{2}\\
\hline
\end{tabular}
\end{table}

Finally, our proposed BDG competes with the former online search algorithm in ``\verb|Pailitao|''. Different from the above settings, we simulate the online environments with multi-shards, and fix the candidate pool size to get satisfactory recall. The comparison is made on the ``others'' set, which is split into fifteen shards to make one shard contain 100M vectors. We merge the results from different shards to attain the final top60 (also based on binary linear search), and show different recalls in Table \ref{tab:pailitao}. It is clear our new method goes beyond the former on matter in accuracy or efficiency.

\section{Conclusions}

Graph-based ANNS algorithms have shown huge advantage in search efficiency and recall. However, most existing methods taking no consideration of offline issue make it far away from being utilized in practice. In this paper, we propose a new approach named {\it Binary Distributed Graph} to solve the problem. By combining the binary code with graph structure and completely implementing the offline graph building in a distributed system, we make a billion-scale graph-construction less than five hours, and performs comparably with the state-of-the-art in online search. Last but not least, we offer a practical and attractive solution in ``\verb|Pailitao|''.

%
% The next two lines define the bibliography style to be used, and the bibliography file.
\bibliographystyle{ACM-Reference-Format}
\bibliography{sample-base}
% we do not need publisher and organization, unify the name of C/J

\end{document}